\documentstyle[12pt]{article}
\def\be{\begin{equation}}
\def\ee{\end{equation}}
\begin{document}
\
\vspace{3cm}
\begin{flushright}
Preprint ITP-96-16E\\
hep-th/9703037
\end{flushright}
\vspace{1cm}
\renewcommand{\thefootnote}{\arabic{footnote}}

\vspace{1cm}
\renewcommand{\thefootnote}{\arabic{footnote}}

\begin{center}
{{\Large \bf
Duality  of the 2D Nonhomogeneous Ising Model}}\\
{{\Large \bf
on   the Torus}}\\
\vspace{1truecm}
{ A.I.~Bugrij
\footnote
{E-mail address:
abugrij\@gluk.apc.org },  V.N.~Shadura}\\
\medskip
 {\it Bogolyubov Institute for Theoretical Physics}\\
 \medskip
 {\it 252143 Kiev-143, Ukraine}
\end{center}
\vspace{2truecm}
\begin{abstract}
\begin{sloppypar}

  Duality relations for the 2D nonhomogeneous
Ising model on the finite square lattice
 wrapped on the torus are obtained.
 The partition function of  the model on
  the dual lattice  with arbitrary combinations of
the periodical and  antiperiodical boundary conditions along
 the cycles of the torus is expressed through
 some specific combination of the partition functions of the model on
 the original  lattice with corresponding boundary conditions.
 It is shown that  the structure of the  duality  relations
 is connected with the topological peculiarities of the dual transformation
 of the model on the torus.

 \end{sloppypar}
\end{abstract}
\thispagestyle{empty}

\newpage
\pagenumbering{arabic}

\medskip
The duality relation for the two-dimensional Ising model was discovered
by Kramers and Wannier [1]. In their work,  Kramers and
Wannier showed the correspondence between the partition function of the
model in low-temperature phase and the partition function of the model on the
dual lattice in high-temperature phase and vice versa:
\be (\cosh
2\widetilde{K}))^{-N}\widetilde{Z}(\widetilde{K})=
(\cosh
2K)^{-N}Z(K)
\ee
$$
\sinh2K\cdot\sinh2\widetilde{K}=1.
$$
 Using this self-duality property,  the critical
 temperature of the 2D Ising model was determined [1] before   Onsager
had obtained the exact solution [2].

Kadanoff and Ceva [3]  generalized   the   Kramers-Wannier duality relation
(1) to the nonhomogeneous case (the coupling constants are
arbitrary functions of lattice site coordinates) with
spherical boundary conditions
 \be \prod_{\widetilde{r},\mu}
\bigl(\cosh 2\widetilde{K}_\mu(\widetilde{r})\bigr)^{-1/2}\widetilde{Z}\bigl[
\widetilde{K}\bigr]=
\prod_{r,\mu} \bigl(\cosh {2}K_\mu(r)\bigr)^{-1/2}Z\bigl[K\bigr],
\ee
\be
\sinh 2K_x(r)\cdot\sinh2\widetilde{K}_y(\widetilde{r})=1,\,\,
\sinh 2K_y(r)\cdot\sinh2\widetilde{K}_x(\widetilde{r})=1.
\ee
Here $\mu=x,y$ and $r$, $\widetilde{r}$,  $K_\mu(r)$,
 $\widetilde{K}_\mu(\widetilde{r})$ are  coordinates and coupling constants on
the original  and dual lattices respectively.  Since the Kadanoff-Ceva relation
(2) defines the connection between functionals, this
relation is very useful for analysis of the thermodymamic phases of the model.
Thus, for example, this relation allows one to  define correctly the disorder
 parameter $\mu$, to obtain the duality relation connecting correlation
 functions on  the original and dual lattices, to define "mixed" correlation
 functions $\langle\sigma(r_i)\dots\sigma(r_j)\mu(r_k)\dots\mu(r_l)\rangle$ and
so on (see Ref. [3]).

As was already mentioned in Ref. [1,3],  relations (1) and (2)
can not be understood  literally.
So, for example, using the method of comparing  high- and
low-temperature expansions for deriving the duality relation (1)
in the case of
 the periodical boundary conditions, it is hard to take into account
 and to compare the graphs
wrapping up the torus. In fact Eq. (1)
is correct in the thermodynamic limit (for the specific free energy).  However
for the nonhomogeneous case the procedure of thermodynamic limit is rather
ambiguous.  In Ref. [3] the duality relation (3) was obtained for spherical
(nonphysical for the square lattice) boundary conditions.

 Since  duality is a popular  method of  nonperturbative investigation in
field theory and statistical mechanics (for review see Ref. [4]),  it is
important to formulate a duality transformation for finite systems.  Recently,
we have suggested [5,6] the exact duality relations for the nonhomogeneous
Ising model on a finite square lattice of size $N=n\times m$ wrapped on the
 torus:
\be
\prod_{\widetilde{r},\mu}\bigl(\cosh2\widetilde{K}_\mu(\widetilde{r})
\bigr)^{-1/2}{\widetilde{\bf Z}}[\widetilde{K}]=
\prod_{r,\mu}\bigl(\cosh2K_\mu(r)\bigr)^{-1/2}\widehat{T}{\bf Z}[K].
\ee
  Here
\be \widehat{T}={1\over 2}\left(\begin{array}{rrrr} \ 1&\ 1&\ 1&\ 1\\
\ 1&\ 1&-1&-1\\ \ 1&-1&\ 1&-1\\ \ 1&-1&-1&\ 1\end{array}\right), \quad \widehat T^2=1.
\ee
and components of
 the  four-vector (${ \widetilde{\bf Z}}[\widetilde K]$ for the dual lattice)
$$
{\bf Z}[K]=(Z^{(p,p)},Z^{(p,a)},Z^{(a,p)},Z^{(a,a)}),
$$
are  partition functions  $Z^{(\alpha,\beta)}[K]$ ($\alpha,\beta = a,p$)
of the Ising model with corresponding combinations of
the periodical $(p)$ and antiperiodical $(a)$ boundary conditions
along  the horizontal $X$ and
vertical $Y$ axes:
\be
Z^{(\alpha,\beta)}[K]=\sum_{[\sigma]}e^{-\beta
{H}^{(\alpha,\beta)}[K, \sigma]},
\ee
\be
-\beta
{H}^{(\alpha,\beta)}[K,\sigma]=
\sum_{r}\sigma(r)\bigl(K_x(r)\nabla^\alpha_x+K_y(r)
\nabla^\beta_y\bigr)\sigma(r),
\ee
where $r=(x,y)$    denotes the site coordinates on the square lattice of
size $N=n\times m$,  $x=1,\dots,n$ $y=1,\dots,m$;  $\sigma(r)=\pm1$;
$K_x(r)$ and $K_y(r)$ are  the coupling constants
along  the horisontal $X$ and
vertical $Y$ axes respectively.
The one-step shift operators  $\nabla_x$, $\nabla_y$
 act on $\sigma(r)$  in the following way
$$
\nabla_x\sigma(r)=\sigma(r+\widehat{x}),\quad\nabla_y\sigma(r)=\sigma
(r+\widehat{y}),
$$
where $\widehat{x}$,  $\widehat{y}$ are  the unit vectors along
the horisontal and vertical axes.
For   the periodical (antiperiodical) boundary conditions along $X$ and $Y$ axes
we have
$$
\nabla_x^{p(a)} \sigma(n,y)=+(-)\sigma(1,y),\quad
\nabla_y^{p(a)} \sigma(x,m)=+(-)\sigma(x,1).
$$
We denote site coordinates, functions and functionals  on
the dual lattice by "tilda" :
$ \widetilde{r}, \,$
$\widetilde{\sigma}(\widetilde{r}), \,$ $ \widetilde{K}_\mu(\widetilde{r}),
\, $ $\widetilde{H}[\widetilde{K},\widetilde{\sigma}], \,$
$\widetilde{Z}[\widetilde{K}], \,\, \dots\, $ .
A site coordinate on the dual lattice coincides with a coordinate
of the plaquet center on  the original lattice:
\be
\widetilde{r}=r+(\widehat{x}+\widehat{y})/2.
\ee

In Ref. [5] the duality relation (4) was proved
for  homogeneous  and
  weakly nonhomogeneous distributions of
the coupling constants.  We also have checked   the duality relation (4) for
lattices of small sizes by direct calculation on the computer.
As a corollary of Eq. (4),   we
obtained [5,6] the duality relations  for the two-point correlation function
on the torus, for the partition functions of  the 2D Ising model  with
magnetic fields applied to the boundaries and   the 2D Ising model with
free, fixed and mixed boundary conditions.

In this Letter we would like to propose a simple way of  the derivation
 of  the duality relation (4).
For the derivation it is convenient to use the representation of
 Hamiltonian (7) in terms of
Hamiltonian $H_D^{(\alpha,\beta)}$ of the Ising model with
the magnetic dislocation corresponding to   the boundary conditions
$(\alpha,\beta)$ and    the periodical boundary conditions for $\nabla_\mu$:
$$
H_D^{(\alpha,\beta)}[K^{(\alpha,\beta)},\sigma] =
\sum_{r,\mu}\sigma(r)K^{(\alpha,\beta)}_\mu (r)\nabla^p_\mu
\sigma(r), \quad \mu=x,y.
$$
 Here  the coupling constants configurations $[K^{(\alpha,\beta)}]$
  define the following
magnetic dislocations:

(i) For  the Hamiltonian   $H_D^{(p,a)}$:
\begin{eqnarray*}
D_ {X}= \left\{\begin{array}{ll}
&K^{(p,a)}_x (r')=K_x (r'), \\ &K^{(p,a)}_y (r')=-K_y (r'),\,\,
\mbox{if}\,\, {r'} \in B^{(m)}_X ; \\
&K^{(p,a)}_\mu (r)=K_\mu (r), \,\,\mbox{if} \,\,r \not\in B^{(m)}_X ; \end{array}\right.
\end{eqnarray*}
where we introduced the denotion $B^{(i)}_X$  for the following boundary cycles
on the torus $$ B^{(i)}_X= \left\{ (x,i), x=1,...,n \right\},\quad i=1,m, $$

(ii) For  the Hamiltonian   $H_D^{(a,p)}$:
\begin{eqnarray*}
D_Y= \left\{\begin{array}{ll}  &K^{(a,p)}_x (r')=-K_x (r'), \\ &K^{(a,p)}_y (r')=K_y (r'),\,\,
\mbox{if}\,\, r'\in B^{(n)}_Y ; \\
&K^{(a,p)}_\mu (r)=K_\mu (r), \,\,\mbox{if} \,\,r
\not\in
B^{(n)}_Y ; \end{array}\right.
\end{eqnarray*}
where  the path $B^{(i)}_Y$  denotes the other boundary cycles  on the torus
$$
  B^{(i)}_Y= \left\{ (i,y), y=1,...,m \right\}, \quad i=1,n,
$$

(iii) For  the Hamiltonian   $H_D^{(a,a)}$ we  have $D_{X,Y}=D_X\cup  D_Y$.

It is evident that  $H_D^{(p,p)}$ has not magnetic dislocation.
Nevertheless, let us  denote
configuration of the coupling constants in this case as
$D_0$
and call   coupling constants configurations $D_0, D_X, D_Y, D_{X,Y} $ as
  the basic magnetic dislocations for corresponding $H_D^{(\alpha,\beta)}$.

Note that  the partition function (6) is invariant with respect to   the
$Z_2$ local gauge transformation $\widehat U [\tau ]$ [7,8]:
\be
\widehat U\bigl[\tau\bigl]\,\bigl[K,\sigma\bigl]=\bigl[{K}^\prime,
\sigma^\prime\,\bigl]=
\bigl[\{\bar{K}_{\mu}(r)=\tau(r){K}_{\mu}(r)\tau(r+\widehat\mu)\},
\,\,\left\{\sigma^\prime(r)=\tau(r)\sigma(r)\right\} \bigl],
\ee
where $\tau(r)=\pm 1$.
Let    us apply arbitrary gauge
transformation to  the partition function $ Z^{(\alpha,\beta)} $
$$
\widehat U\bigl[\tau\bigl]
\sum_{[\sigma]}\exp\{{-\beta H_D^{(\alpha,\beta)}[K, \sigma]}\} =
\sum_{[\sigma^\prime]}\exp\{{-\beta H_D^{(\alpha,\beta)}[{K}^\prime,
\sigma^\prime]}\} = Z^{(\alpha,\beta)}\bigl[{K}^\prime\bigl]
=Z^{(\alpha,\beta)}[K].
$$
Here new coupling constants configuration $\bigl[{K}^\prime\bigl]$
 can contain both the deformation of the corresponding basic
magnetic dislocations and new closed magnetic dislocations.  Let us denote by
$\Omega^{(p,p)},\Omega^{(p,a)},\Omega^{(a,p)},\Omega^{(a,a)}$  the classes of
 gauge
equivalent  configurations $\bigl[{K}^\prime\bigl]$ generated by  the gauge
 transformations from  the basic magnetic dislocations $D_0, D_X, D_Y,
D_{X,Y} $ respectively.  It is evident that these classes have not
intersections  because
the basic magnetic dislocations are the homotopy-nonequivalent paths on the
two-dimensional torus (two arbitrary coupling constants configurations from
different classes $\Omega^{(\alpha,\beta)}$ can't be connected by
the $Z_2$  gauge transformations).

For the dual transformation of  the partition function  (6) let us  use
 the standard method [4] of the passage to  dual spins. At the
beginning we consider   the dual transformation of $Z^{(p,p)}$:
$$
Z^{(p,p)}[{K}]= \sum_{[\sigma]} \exp \bigl[ \sum_{r,\mu}\sigma(r)K_\mu
(r)\nabla^p_\mu \sigma(r)\bigl]= $$ $$ \sum_{[\sigma]} \prod_{r,\mu} \bigl[
\sum_{l_\mu}P_{l_\mu(r)}(K_\mu(r))(\sigma(r)
\sigma(r+\widehat\mu))^{l_\mu(r)}\bigl]=
 $$
\be
\sum_{[l_\mu]}
\sum_{[\sigma]} \prod_{r,\mu}
P_{l_\mu(r)}(K_\mu(r))\prod_{r}(\sigma(r))^{\psi(r)}=
2^N \sum_{[l_\mu]}
 \prod_{r,\mu}
P_{l_\mu(r)}(K_\mu(r))\prod_{r}\delta_2[\psi(r)]
\ee
where
\be
P_{l_\mu(r)}(K_\mu(r))=\cosh(K_\mu(r))\exp(l_\mu(r)\ln\tanh K_\mu(r))
\ee
and
\be
\psi(r)=l_x(r)+l_y(r)+ l_x(r-\widehat x)+l_y(r-\widehat y),\quad l_\mu(r)=0,1.
\ee
Here $\delta_2(\psi)$ is a Kronecker $\delta$-function   $mod \,2$ :
it is zero if $\psi$ is odd and one if  $\psi$   is even.
Note that in Eq. (10) we have a product  of the $\delta$-functions with linking
arguments (the same $l_\mu$ is contained in two $\delta$-functions).  In order
to solve constraints generated by the product of $\delta$-functions
$l_\mu(r)$ is usually expressed through dual spins $\widetilde{\sigma}
(\widetilde r)$ [4]
\be l_\mu(r) ={1\over 2} (1-\widetilde{\sigma}(\widetilde r) \widetilde{\sigma}
(\widetilde r-\nu)),\quad  \mu \neq \nu .
\ee
Substituting Eq. (12) in Eq. (9) we
obtain $$ Z^{(p,p)}[{K}] ={1\over 2} \sum_{[\widetilde\sigma]}
\prod_{r,\mu}(\sinh 2K_\mu(r))^{1\over 2} \exp\bigl[\sum_{\widetilde
r,\nu}\widetilde\sigma(\widetilde r) \widetilde K_\nu (\widetilde
r)\nabla^p_\nu \widetilde\sigma(\widetilde r)\bigl]=
$$
\be
{1\over 2}
\prod_{\widetilde r,\mu}(\cosh 2\widetilde K_\nu(\widetilde r) )^{-{1}}
{\widetilde Z}^{(p,p)}[{\widetilde K}],
\ee
where the dual coupling constants
$\widetilde K_\nu(\widetilde r)$ are defined by (3) or
\be \tanh K_\mu(r) =
  e^{-2\widetilde K_\nu(\widetilde r)},\quad \mu \neq \nu.
\ee
To derive  Eq. (14) we
 used   identity
$$ {\cosh^2 2K_\mu(r)\over\sinh2K_\mu(r)}=
{\cosh^2 2\widetilde{K}_\nu(\widetilde{r})\over\sinh2\widetilde{K}_\nu
(\widetilde{r})}, \quad \mu \neq \nu.
$$
Since in  Eq. (14) we sum over $[\widetilde\sigma]$ (this counts each
 configuration $ [l_\mu]$ twice because  $[\widetilde\sigma] \rightarrow
 [-\widetilde\sigma]$ gives the same $ [l_\mu]$),  we must introduce factor
$1/2$.

However it is not hard to note  that we can construct many other
solutions for $l_\mu$
for which the coupling constant configurations are connected
with $[{\widetilde K}]$ in (14) by means of the gauge transformation
$\widehat {\widetilde U}\bigl[\widetilde \tau \bigl]$ on the dual lattice (see
Eq. (9)).
Such   configurations form a class $\widetilde \Omega^{p,p}$
of gauge-equivalent coupling constant configurations .
By analogy with the Ising model on the original lattice wrapped on the torus
we must expect the existence of  solutions for $l_\mu$ which lead to
homotopy-nonequivalent classes $\widetilde\Omega^{(p,p)},
\widetilde\Omega^{(p,a)},\widetilde\Omega^{(a,p)},\widetilde\Omega^{(a,a)}$
of dual coupling constants configurations.
Really it is easy to write the solutions for $l_\mu$  which
lead to the basic  magnetic dislocations
${\widetilde D}_X,
{\widetilde D}_Y, {\widetilde D}_{X,Y} $
in $\widetilde\Omega^{(p,a)},\widetilde\Omega^{(a,p)},\widetilde\Omega^{(a,a)}$
respectively.
So, taking into account Eq. (8),  we have for ${\widetilde D}_X$:
\be
l_x(r)= {1\over 2} (1+\widetilde{\sigma}(\widetilde r)
\widetilde{\sigma} (\widetilde r-\widehat y)),
\ee
$$
l_y(r)= {1\over 2} (1-\widetilde{\sigma}(\widetilde r)
\widetilde{\sigma} (\widetilde r- \widehat x)),\,\,
\mbox{if} \,\,{r} \in B^{(1)}_X ;
$$
$$
l_\mu(r)={1\over 2} (1-\widetilde{\sigma}(\widetilde r)
\widetilde{\sigma} (\widetilde r-\widehat\nu)),\,\,
\mbox{if} \,\,r \not\in B^{(1)}_X  ;
$$
for ${\widetilde D}_Y$:
\be
l_y(r)= {1\over 2} (1+\widetilde{\sigma}(\widetilde r)
\widetilde{\sigma} (\widetilde r-\widehat x)),\,\,
\ee
$$
l_x(r)= {1\over 2} (1-\widetilde{\sigma}(\widetilde r)
\widetilde{\sigma} (\widetilde r-\widehat y)),\,\,
\mbox{if} \,\,{r} \in B^{(1)}_Y  ;
$$
$$
l_\mu(r)={1\over 2} (1-\widetilde{\sigma}(\widetilde r)
\widetilde{\sigma} (\widetilde r-\nu)),\,\,
\mbox{if} \,\,r \not\in B^{(1)}_Y  ;
$$
and for ${\widetilde D}_{X,Y}$:
\be
l_x(r)= {1\over 2} (1+\widetilde{\sigma}(\widetilde r)
\widetilde{\sigma} (\widetilde r-\widehat y)),\,\,
\ee
$$
l_y(r)= {1\over 2} (1-\widetilde{\sigma}(\widetilde r)
\widetilde{\sigma} (\widetilde r-\widehat x)),\,\,
\mbox{if} \,\,r \in B^{(1)}_X  ;
$$
$$
l_y(r)= {1\over 2} (1+\widetilde{\sigma}(\widetilde r)
\widetilde{\sigma} (\widetilde r-\widehat x)),\,\,
$$
$$
l_x(r)= {1\over 2} (1-\widetilde{\sigma}(\widetilde r)
\widetilde{\sigma} (\widetilde r-\widehat y)),\,\,
\mbox{if} \,\,r \in B^{(1)}_Y  ;
$$
$$
l_\mu(r)={1\over 2} (1-\widetilde{\sigma}(\widetilde r)
\widetilde{\sigma} (\widetilde r-\nu)),\,\,
\mbox{if} \,\,r \not\in B^{(1)}_X , B^{(1)}_Y .
$$
In Eqs. (16)-(18) $ \mu \neq \nu $.

Since arbitrary solutions satisfying the product of $\delta$-functions in
Eq. (10) lead to the dual coupling constant configurations which contain the
finite number of closed magnetic dislocations on the dual lattice, it is
obvious that these configurations can be generated by the gauge transformations
from the basic magnetic dislocations and occur in the corresponding  class
$\widetilde\Omega^{(\alpha,\beta)}$.
Then in Eqs. (10), (14) at transformation to the dual lattice  we must sum
over  the homotopy-nonequivalent solutions (13), (16)-(18).
Taking into account that  basic magnetic dislocations ${\widetilde D}_0,
{\widetilde D}_X, {\widetilde D}_Y, {\widetilde D}_{X,Y} $ lead to the
partition functions with corresponding boundary conditions,
we obtain:
$$
\prod_{r,\mu}\bigl(\cosh 2K_\mu(r)\bigr)^{-{1\over 2}}Z^{(p,p)}[{K}] =
$$
\be
{1\over
2}\prod_{\widetilde{r},\nu}\bigl(\cosh 2\widetilde{K}_\nu(\widetilde{r})
\bigr)^{-{1\over 2}}\left({\widetilde{ Z}^{(p,p)}}[\widetilde{K}]+
{\widetilde{ Z}^{(p,a)}}[\widetilde{K}]+
{\widetilde{ Z}^{(a,p)}}[\widetilde{K}]+
{\widetilde{ Z}^{(a,a)}}[\widetilde{K}]\right).
\ee
Now, let us consider the dual transformation of  partition
function $Z^{(p,a)}$. As we have discussed above, it contains
magnetic dislocation $D_X$.
Again, doing the duality transformation,
it is necessary to sum over  the homotopy-nonequivalent solutions (13),
(16)-(18). As a result we obtain the duality relations similar to Eq. (19) but with the
minus signs before  ${\widetilde{ Z}^{(p,a)}}[\widetilde{K}]$ and
${\widetilde{Z}^{(a,a)}}[\widetilde{K}]$ .
Appearence of these signs is connected with  the
presence of magnetic dislocation $D_X$ in $Z^{(p,a)}$.  Let us show this.
Considering the transformation to the dual lattice in the same way as in Eqs.
(10), (14),  we obtain  the partition function
${\widetilde{ Z}^{(p,p)}}\bigl[\widetilde{K}^{(p,p)}\bigl]$
with
the following dislocation (see Ref. [3]):
\begin{eqnarray*}
\widetilde G_X= \left\{\begin{array}{ll}
 &{{\widetilde K}^{(p,p)}}_x ({{\widetilde r}^\prime}))={\widetilde K}_x (
 {{\widetilde r}^\prime})+i{\pi\over 2},
\\ &{{\widetilde K}^{(p,p)}}_y
({\widetilde r}^\prime)={\widetilde K}_y ({{\widetilde r}^\prime}),\,\,
\mbox{if} \,\,{\widetilde r}^\prime
\in {\widetilde B}^{(m)}_X;
\\ &{{\widetilde K }^{(p,p)}}_\mu
({{\widetilde r}})={\widetilde K}_\mu (\widetilde r),\,\, \mbox{if}
\,\,\widetilde r \not\in {\widetilde B}^{(m)}_X ; \end{array}\right.
 \end{eqnarray*}
where the  path
${\widetilde B}^{(m)}_X$ denotes the following boundary cycle on
the dual torus
 $$ {\widetilde B}^{(m)}_X= \left\{ (\widetilde x, m), \widetilde x=1,...,n
 \right\}, $$
that is  the dual transformation transforms dislocation  ${ D}_X$
  to dislocation $\widetilde G_X$.

The partition function
${\widetilde{ Z}^{(p,p)}}\bigl[\widetilde{K}^{(p,p)}\bigl]$
can be  written through
the following correlation function [3]
 $$
 i^n<\widetilde{\sigma}(1,m)\widetilde{\sigma}(2,m)\widetilde{\sigma}(2,m)
 \widetilde{\sigma}(3,m)...\widetilde{\sigma}(n,m)\widetilde{\sigma}(n+1,m)>
 {\widetilde{ Z}^{(p,p)}}[\widetilde{K}]=i^n
{\widetilde{ Z}^{(p,p)}}[\widetilde{K}],
 $$
where  ${\widetilde{ Z}^{(p,p)}}[\widetilde{K}]$ does not contain magnetic
dislocations.  In other hand, observing that
except the dislocation $\widetilde G_X$ the partition functions
 ${\widetilde{Z}^{(p,a)}}\bigl[\widetilde{K}^{(p,a)}\bigl]$ and ${\widetilde{
 Z}^{(a,a)}}\bigl[\widetilde{K}^{(a,a)}\bigl]$
have
magnetic dislocation ${\widetilde D}_y$, we obtain, for example, for
the last partition function
$$
i^n<\widetilde{\sigma}(1,m)\widetilde{\sigma}(2,m)\widetilde{\sigma}(2,m)
 \widetilde{\sigma}(3,m)...\widetilde{\sigma}(n,m)\widetilde{\sigma}(n+1,m)>
 {\widetilde{ Z}^{(a,a)}}[\widetilde{K}]=-i^n{\widetilde{Z}^{(a,a)}}
[\widetilde{K}].
 $$
 Here the  minus sign appeares because of the  antiperiodical  boundary
conditions.  Similarly the minus sign appears before
 ${\widetilde{Z}^{(p,a)}}\bigl[\widetilde{K}\bigl]$.  Then, taking into account  these
signs, one gets
$$
\prod_{r,\mu}\bigl(\cosh2K_\mu(r)\bigr)^{-{1\over 2}}Z^{(p,a)}[{K}]
=
$$
\be
{1\over 2}\prod_{\widetilde{r},\nu}\bigl(\cosh2\widetilde{K}_\nu(\widetilde{r})
\bigr)^{-{1\over 2}}\left({\widetilde{ Z}^{(p,p)}}[\widetilde{K}]-
{\widetilde{ Z}^{(p,a)}}[\widetilde{K}]+
{\widetilde{ Z}^{(a,p)}}[\widetilde{K}]-
{\widetilde{ Z}^{(a,a)}}[\widetilde{K}]\right),
\ee
 where the factor $i^n$ is cancelled by the same factor appearing
 from the normalizing product  $\prod \cosh^{1/2}\widetilde{K}$.
The same way, considering  the dual transformation for $Z^{(a,p)}$
and $Z^{(a,a)}$,  we obtain
\be
\prod_{{r},\mu}\bigl(\cosh2{K}_\mu({r})
\bigr)^{-1/2}{{\bf Z}}[{K}]=
\prod_{\widetilde r,\mu}\bigl(\cosh2\widetilde K_\mu(\widetilde r)
\bigr)^{-1/2}\widehat{T}\widetilde{\bf Z}[\widetilde K].
\ee
 Multiplying  the right and the left sides of Eq. (21) by matrix
 $\widehat T $  (5),  one obtains Eq. (4).

 Thus the structure of duality relation (4) is connected with the topological
 peculiarities of the dual transformation of the Ising model on the torus.
 From here, one can assume that dual relations for  other models with $Z_2$
symmetry, for example, the eight-vertex model, have the similar structure.  The
 method of derivation of duality relation (4) suggested in this paper
 can be generalized for the two dimensional lattice models with $Z_N$ symmetry.
 Results of this reseach will be published in the following paper.

The authors are indebted to A.~Belavin and A.~Morozov for continuous support
and are grateful   to M.~Lashkevich and A.~Mironov for useful discussions.
One of the authors (V.S.) thanks  A.~Morozov for the hospitality and the
exellent conditions at ITEP  where this work has been begun.
This work is
partially supported by grant INTAS-93-1038.

\medskip

\centerline{\bf Reference}

\medskip

\begin{enumerate}
\item H.A.~Kramers, G.H.~Wannier, Phys. Rev.
{\bf 60}, 252 (1941).
\item L.~Onsager, Phys. Rev. {\bf 65}, 117
(1944).
\item L.P.~Kadanoff, H.~Ceva, Phys. Rev. B {\bf 3}, 3918
(1971).
\item R.~Savit, Rev. Mod. Phys. {\bf 52}, 453 (1980).
\item A.I.~Bugrij, V.N.~Shadura, Zh. Eksp. Teor. Fiz.
 {\bf 109}, 1024 (1996).
\item  A.I.~Bugrij, V.N.~Shadura, Pis'ma  Zh. Eksp. Teor. Fiz.
 {\bf 63}, 369 (1996).
\item G.~Toulouse, Commun. Phys. {\bf 2}, 115 (1977).
\item E.~Fradkin, V.A.~Huberman, S.H.~Shenker, Phys. Rev.
 B {\bf 18}, 4789 (1978).
 \end{enumerate}

\end{document}